\documentclass{aa}  
\usepackage{graphicx}
\usepackage{txfonts}

\def\micron{\hbox{$\mu$m}}

\begin{document}
   \title{A VLA search for young protostars embedded in dense cores}

   \author{D. Stamatellos\inst{1}
          \and
          D. Ward-Thompson\inst{1}$^,$\inst{2}
	  \and
          A.~P. Whitworth\inst{1}
	  \and 
	  S. Bontemps\inst{2}	
          }

    \offprints{D. Stamatellos \\ \email{D.Stamatellos@astro.cf.ac.uk}}

\institute{School of Physics \& Astronomy, Cardiff University, 5 The Parade, Cardiff, CF24 3AA, Wales, UK
\and	Observatoire de Bordeaux, BP 89, 2 Rue de l'Observatoire, 33270 Floirac, France}

 \date{Received 13 September, 2006; accepted 6 November 2006}

 
\abstract
{}
{Four dense cores, L1582A, L1689A, B133 and B68, classified as prestellar
in  terms of the absence of detectable NIR emission, are observed at radio wavelengths 
to investigate whether they nurture very young protostars.}
{We perform deep radio continuum observations at 3.6 cm and 6 cm using the 
VLA.}
{No definite young protostars were discovered in any of the four cores observed. A few radio
sources were discovered close to the observed cores, 
but these are most likely extragalactic sources or YSOs unrelated to the cores 
observed.

In L1582A we discovered a weak radio source near the centre of the core with
radio characteristics and offset from the peak of the submillimeter
emission similar to that of the newly discovered protostar in the core L1014,
indicating a possible protostellar nature for this source.  This needs to be confirmed 
with near- and/or mid-infrared observations 
(e.g. with {\it Spitzer}). 
Hence based on the current observations we are unable to confirm unequivocally that
L1582A is starless.

In L1689A a possible 4.5-$\sigma$ radio source was discovered at the centre of the core,
but needs to be confirmed with future observations.

In B133 a weak radio source, possibly a protostar, was discovered at the
edge of the core on a local peak of the core submm
emission, but no source was detected at the centre of the core. Thus, B133
is probably starless, but may have a protostar at its edge.

In B68 no radio sources were discovered inside or at the edge of the core,
and thus B68 is indeed starless.

Four more radio sources with spectral indices characteristic of 
young protostars were discovered outside the cores but within the 
extended clouds in which these cores reside.}
{We conclude that the number of cores misclassified as prestellar is probably very small
and does not significantly alter the  estimated lifetime of the prestellar phase. }


   \maketitle
%

\section{Introduction}

The identification of the youngest protostars is important for
understanding how dense cores in molecular clouds collapse and form stars.
The initial conditions for star formation are provided by dense cores in
molecular clouds in which there is no evidence that star formation has
occurred (starless cores; Myers et al. 1983). Some of these cores are
thought to be close to collapse or already collapsing, and they are
labelled prestellar cores (e.g. Ward-Thompson et al. 1994;  Ward-Thompson
et al. 2002). Prestellar cores evolve to Class 0 objects, which represent
the youngest protostars (Andr\'e et al. 1993).

The presence of a protostar in a core is inferred by one or more of the
following criteria (Andr\'e et al. 2000): (i) compact centimeter radio
emission, (ii) a bipolar molecular outflow, (iii) NIR or MIR emission.
These criteria depend on the sensitivity of the telescope used, and how
young and embedded the protostar is in its parent cloud.
Additionally, the amount of radio emission from young protostars is
uncertain and probably very low (e.g. Curiel et al. 1987; Rodr\'iguez et al.
1989b; Neufeld \& Hollenbach 1996; Harvey et al. 2002;
Eiroa et al. 2005).  Finally, in many cases the complexity of star forming
regions (e.g. dense environment, multiple sources, inflow of material) 
does not allow a clear identification of all the outflows in a
given region.  Thus, a null detection with a particular telescope does
not necessarily mean that 
a source does not exist inside a core.

The need for very sensitive observations to discover young protostars residing
inside dense cores was demonstrated recently by Young et al. (2004), who, 
using the {\sc {\it Spitzer}} space telescope, detected a NIR source 
embedded in L1014 (L1014-IRS), a dense core that was previously classified
as starless.  This suggests that L1014 is in fact a young Class 0 object.
Since the Young et al.  discovery, the protostellar nature of L1014-IRS
has been confirmed by the detection of a very weak bipolar outflow
emanating from the IR source (Bourke et al. 2005) that affects the geometry of
the observed scattering light nebula around the source (Huard et al.
2006).  Additionally 2 more NIR sources have been discovered inside cores
previously thought to be starless: in L1521F in the Taurus molecular cloud
(Terebey et al. 2005; Bourke et al. 2006), 
and in L1148 in the Cepheus flare (Kauffmann et al.
2005). These recently discovered low-luminosity protostars are
similar to the previously known Class 0 protostars VLA1623 (Andr\'e et al. 1993)
and  HH24MMS (Ward-Thompson et al. 1995; Bontemps et al. 1996).
These discoveries of low-luminosity, young protostars,
open up the possibility that other cores
previously classified as prestellar may in fact contain very young
protostars.

Radio observations are suitable for detecting deeply embedded protostars,
as long wavelength radiation can escape the large column density of the 
envelope practically
unattenuated. Previous studies (e.g. Andr\'e et al. 1987; Leous et al.
1991; Bontemps et al. 1995) have identified a large number of young
protostars exhibiting detectable radio emission. However, very young and
low luminosity protostars are believed to produce very weak radio emission
(e.g. Harvey et al. 2002; Neufeld \& Hollenbach 1994, 1996), and thus deep
observations are needed to detect them.

Young protostars exhibit a positive or almost flat ($>-0.1$)
spectral index (defined as $\alpha$, where $S_\nu\propto \nu^a$) 
between e.g. 3.6 cm and 6 cm
(Andr\'e et al.1987; Anglada et al. 1998; Beltr\'an et al. 2001). This
behaviour of the radio spectral index is characteristic of free-free
thermal emission from ionised gas. This is most likely produced by a
partially ionised jet that propagates into the collapsing envelope
producing shocks (e.g. Curiel et al. 1987).  It could also be produced by
an ionised disc wind (e.g. Martin 1996) or by the accretion shock that
develops on the surface of the protostar, and/or on the surface of the
disc that surrounds the protostar, which heats and ionises the infalling
gas (Winkler \& Newman 1980; Cassen \& Moosman 1981). However, the latter
two mechanisms produce radio emission that is at too low a level to be
readily detected. 

A negative ($<-0.1$) radio spectral index indicative of
non-thermal gyro-synchrotron emission has also been observed in some T
Tauri stars  (e.g. Andr\'e et al. 1988;  Andr\'e 1996; Rodr\'iguez et al. 1999), 
in a possible Class I object (Feigelson et al. 1998), and 
even in a possible Class 0 object, the triple
radio continuum source in Serpens (Rodr\'iguez et al. 1989a;
Curiel et al. 1993; Curiel 1995;
Raga et al. 2000). The non-thermal emission of more evolved
protostars (Class II, III) is believed to be associated 
with the magnetic fields around these objects and the star-disc interaction region
(e.g. Andr\'e 1996). The non-thermal emission from younger protostars
is thought to be the result of particle acceleration behind a diffuse shock wave
(Rodr\'iguez et al. 1989a; Curiel et al. 1993 and references therein).

Stamatellos et al. (2005a,b) performed 3-dimensional radiative transfer
calculations on simulations of the collapse of a star-forming core and
studied the transition from a starless core to a core with an embedded
young protostar. These simulations suggest (i) that cores with young
protostars appear to be cold ($T\sim15-30$~K) but still hotter than
starless cores ($T<15$~K), and (ii) that cores with young protostars appear
more circular (at least in their central regions) than starless cores (see
also Goodwin et al. 2002).  Based on this study, we identified 4 cores
classified as starless (L1582A, L1689A, B133, B68) that might
in fact contain very young protostars. These cores were selected in terms
of their circular appearance and their relatively high estimated
temperatures ($\sim15$~K).  These
cores were observed with the VLA in the continuum at 3.6 cm and 6~cm.

In this paper we present the results of these radio observations aiming to
detect very young embedded sources in dense cores.  In Section~2 we
describe the observational details and the properties of the detected
radio sources, and in Section~3 we summarize our results.


\section{Observations and results}

L1582A, L1689A, B133 and B68 were observed at 3.6 cm using the VLA. The
observations were performed in July/August 2005, in the VLA C
configuration.  The total integration time was about $\sim 4$ hours for
each core including time spent on the calibrators. These cores (except
B68) were also observed at 6 cm in October/November 2005, in the DnC
configuration. L1582A was observed for $\sim 5$ hours, B133 for $\sim 5$
hours, and L1689A for $\sim 1$ hour. The data were calibrated using the
AIPS package of the NRAO.

\begin{table}[b]
\begin{minipage}[t]{0.47\textwidth}
\caption{Distances and positions of the dense cores observed}    
\label{tab:cores0}      
\renewcommand{\footnoterule}{}  
\centering             
\begin{tabular}{l c c c}
\hline\hline        
\noalign{\smallskip}
Core &$d$ \footnote {Distances taken from Kirk et al. (2005)}
 (pc) &R.A. \footnote{Peak of the 450~$\micron$ \& 850~$\micron$ emission (Kirk et al. 2005)}
  (2000) & Dec. $^b$ (2000)  \\
\noalign{\smallskip}\hline                        
\noalign{\smallskip}
L1582A &400&  $05^{\rm h}\ 32^{\rm m}\ 01.0^{\rm s}$ & $+12\degr\ 30^\prime\ 23^{\prime\prime}$\\
\noalign{\smallskip}
L1689A &130& $16^{\rm h}\ 32^{\rm m}\ 13.2^{\rm s}$ & $-25\degr\ 03^\prime\ 45^{\prime\prime}$\\
\noalign{\smallskip}
B133   &200& $19^{\rm h}\ 06^{\rm m}\ 08.4^{\rm s}$ & $-06\degr\ 52^\prime\ 52^{\prime\prime}$\\
\noalign{\smallskip}
B68    &130& $17^{\rm h}\ 22^{\rm m}\ 39.2^{\rm s}$ & $-23\degr\ 50^\prime\ 01^{\prime\prime}$\\
\hline                        
\end{tabular}
\end{minipage}
\end{table}

\begin{table*}
\begin{minipage}[b]{\textwidth}
\caption{Phase calibrators, beam sizes, RMS noise, and number of radio sources detected}    
\label{tab:cores}      
\renewcommand{\footnoterule}{}  
\centering             
\begin{tabular}{l c ccc cccc c c ccc}
\hline\hline        
\noalign{\smallskip}
Core &
\multicolumn{3}{c}{Phase Calibrator(s)}  &&  
\multicolumn{2}{c}{Beam (\arcsec)}&&
\multicolumn{2}{c}{RMS noise ($\mu{\rm Jy/bm}$)
\footnote{Mean RMS noise in the field in $\mu{\rm Jy/beam}$}}&
 &\multicolumn{2}{c}{Radio sources\footnote{Number of radio sources detected in each field. 
In the parentheses
is the expected number of background sources calculated using 
Eqs.~\ref{eq:back1},~\ref{eq:back2}.}}\\    
\noalign{\smallskip}
\cline{2-4}\cline{6-7}\cline{9-10}\cline{12-13}
\noalign{\smallskip}
& Calibrator &$S_{\rm 3.6\ cm}$(${\rm Jy}$) &$S_{\rm 6\ cm}$ (${\rm Jy}$)& 
&3.6 cm &6 cm & & 3.6 cm & 6 cm  && 3.6 cm & 6 cm  \\
\noalign{\smallskip}\hline                        
\noalign{\smallskip}
L1582A &
$0530+135$& $2.905\pm0.002$ & $3.08\pm0.09$&&
$3.3\times2.7$&$12.4\times6.6$ &&13 & 29 &&4 (2.1)&2 (2.7) \\
\noalign{\smallskip}
L1689A &
$1522-275$&$1.210\pm 0.006$ &$1.06\pm0.07$  &&
$5.5\times2.9$&$17.1\times 10.4$ &&12 & 60&&6 (2.3) &5 (1.3) \\
&$1626-298$&$2.602\pm0.015$&-&&&&&&\\
\noalign{\smallskip}
B133   &
$1939-100$ & $0.566\pm0.002$ &- && 
$3.7\times3.0$&$16.9\times9.6$ &&11 & 60 &&4 (2.3) &2 (1.3)\\
&$1939-154$&-& $0.55\pm0.01$&&&\\
\noalign{\smallskip}
B68    &
$1743-309$ &$0.233\pm0.001$ &-&& 
$5.3\times3.0$& - &&12 & -&&4 (2.3)&- \\
&$1751-253$& 0.259$\pm0.001$ &-&&&\\
\hline                        
\end{tabular}
\end{minipage}
\end{table*}
\begin{table*}
\begin{minipage}[t]{\textwidth}
\caption{Radio sources detected at 3.6 and/or 6cm}    
\label{tab:radiosources}      
\centering             
\renewcommand{\footnoterule}{}  
\begin{tabular}{lcccccccc}
\hline\hline        
\noalign{\smallskip}
Source &\multicolumn{2}{c}{Position 
\footnote{Positional errors are expected to be $~\sim1\arcsec$}
} &
 $S_{\rm 3.6\ cm}$\footnote{Flux density ($\mu{\rm Jy}$) and 1-$\sigma$ errors,
corrected for the primary beam response.}  
& $S^{\rm peak}_{\rm 3.6\ cm}$\footnote{Peak flux density ($\mu{\rm Jy/beam}$)
and 1-$\sigma$ errors,
corrected for the primary beam response.}&
 $S_{\rm 6\ cm}$ & $S^{\rm peak}_{\rm 6\ cm}$
& $\alpha$\footnote{Spectral index between 3.6 and 6 cm}
& Offset\footnote{Angular displacement of the source from the peak of the submm emission} 
 \\
 & RA (2000) & Dec. (2000) &($\mu{\rm Jy}$)  & ($\mu{\rm Jy/beam}$) & 
($\mu{\rm Jy}$)& ($\mu{\rm Jy}$/beam)& & (\arcsec)\\
\noalign{\smallskip}
\hline                        
\noalign{\smallskip}
L1582A & & & & & &\\
1  \ldots\dots& 
$05^{\rm h}\ 32^{\rm m}\ 00.5^{\rm s}$ & $+12\degr\ 30^\prime\ 40^{\prime\prime}$ & $142\pm23$ &$139\pm13$& $280\pm530$ & $280\pm30$ &$-1.3\pm0.5$ & 18 \\
2  \ldots\dots& 
$05^{\rm h}\ 32^{\rm m}\ 06.3^{\rm s}$ & $+12\degr\ 32^\prime\ 41^{\prime\prime}$&$200\pm60$&$137\pm26$ &-&-&- & 165 \\
3  \ldots\dots&
$05^{\rm h}\ 32^{\rm m}\ 08.9^{\rm s}$ &  $+12\degr\ 29^\prime\ 34^{\prime\prime}$ &-&  - & $300\pm100$ &$180\pm40$ &-& 138\\
L1689A & & & & & & & &\\
1  \ldots\dots&
 $16^{\rm h}\ 32^{\rm m}\ 23.3^{\rm s}$ & $-25\degr\ 00^\prime\ 34^{\prime\prime} $ & $8610\pm130$& $8180\pm70$ & $5800\pm600$  & $ 4600\pm 500$ & $\ \ 0.8\pm0.2$& 235\\
2  \ldots\dots&
 $16^{\rm h}\ 32^{\rm m}\ 25.4^{\rm s}$ & $-25\degr\ 02^\prime\ 16^{\prime\prime} $  &$830\pm70$& $490\pm30$    & $1100\pm400$  &$ 640\pm 170$ & $-0.6\ \ \pm0.7$& 188\\
3  \ldots\dots& 
$16^{\rm h}\ 32^{\rm m}\ 13.9^{\rm s}$ & $-25\degr\ 02^\prime\ 25^{\prime\prime} $  &$544\pm26$& $496\pm14$     & $750\pm70$ & $ 740\pm 70$ &$-0.63\pm0.21$ & 80\\
4  \ldots\dots&
 $16^{\rm h}\ 32^{\rm m}\ 14.2^{\rm s}$ & $-25\degr\ 02^\prime\ 04^{\prime\prime} $  &$520\pm30$&  $347\pm15$   & $550\pm70$ & $ 550\pm 70$ & $-0.11\  \ \pm0.28$&  101 \\
5  \ldots\dots&
 $16^{\rm h}\ 32^{\rm m}\ 13.8^{\rm s}$ & $-25\degr\ 02^\prime\ 38^{\prime\prime} $  &$440\pm40$& $174\pm13$    & $660\pm60$& $ 640\pm 60$ &$-0.79 \pm0.25$&  67\\
6  \ldots\dots& 
$16^{\rm h}\ 32^{\rm m}\ 13.3^{\rm s}$ & $-25\degr\ 03^\prime\ 45^{\prime\prime} $  &$58\pm12$& $58\pm12$       &- &-&- & 1\\
B133 & & & &  & & \\
1 \ldots\dots & 
$19^{\rm h}\ 06^{\rm m}\ 05.7^{\rm s}$ & $-06\degr\ 50^\prime\ 06^{\prime\prime} $& $31850\pm50$ & $28730\pm30$ &$67800\pm300$ &$66550\pm150$& $-1.48\pm0.01$& 171\\
2 \ldots\dots & 
$19^{\rm h}\ 06^{\rm m}\ 14.0^{\rm s}$ & $-06\degr\ 49^\prime\ 35^{\prime\prime} $& $730\pm130$&$420\pm50$&- &-&$>1.7\pm0.3$& 214\\
3 \ldots\dots & 
$19^{\rm h}\ 06^{\rm m}\ 08.4^{\rm s}$ & $-06\degr\ 53^\prime\ 32^{\prime\prime} $& $120\pm30$&$71\pm12$&- &- &-&40\\
4 \ldots\dots & 
$19^{\rm h}\ 06^{\rm m}\ 07.0^{\rm s}$ & $-06\degr\ 53^\prime\ 58^{\prime\prime} $& $100\pm25$&$81\pm13$&- &-&-&69\\
B68\footnote{B68 was not observed at 6 cm} & & & & & & &\\
1   \ldots\dots  & 
$17^{\rm h}\ 22^{\rm m}\ 35.0^{\rm s}$ & $-23\degr\ 47^\prime\ 58^{\prime\prime} $& $9310\pm40$ &$8550\pm21$ &  - & -  & - &136\\
2   \ldots\dots  & 
$17^{\rm h}\ 22^{\rm m}\ 31.8^{\rm s}$ & $-23\degr\ 48^\prime\ 18^{\prime\prime} $& $223\pm24$&$223\pm24$ &  - & -&- & 145\\
3   \ldots\dots  & 
$17^{\rm h}\ 22^{\rm m}\ 28.0^{\rm s}$ & $-23\degr\ 50^\prime\ 32^{\prime\prime} $& $410\pm26$&$410\pm26$ &   - &-&- & 157\\
4   \ldots\dots  & 
$17^{\rm h}\ 22^{\rm m}\ 35.4^{\rm s}$ & $-23\degr\ 52^\prime\ 34^{\prime\prime} $& $150\pm50$& $143\pm27$ &   - &-& - & 166\\
\noalign{\smallskip}
\hline                        
\end{tabular}
\end{minipage}
\end{table*}

In Table~\ref{tab:cores0} we list the positions of the cores taken from
Kirk et al. (2005).  These positions correspond to the peak of the
850~$\micron$ and 450~$\micron$ emission detected using SCUBA with a beam
size FWHM of 14.8\arcsec. { We use these positions as phase centres.}
In Table~\ref{tab:cores} we list the phase calibrators used, their
bootstrapped flux densities, the VLA beam size, and the achieved RMS noise for each field.  
Finally, we report the number of radio sources detected and the number of
background, extra-galactic sources expected in each field (see below).  
{The amplitude calibrators used were 
3C286 ($F_{\rm 3.6 cm}=5.26$~Jy,  $F_{\rm 3.6 cm}=7.34$~Jy)
and 3C48 ($F_{\rm 3.6 cm}=3.25$~Jy,  $F_{\rm 3.6 cm}=5.32$~Jy).}

The number of background sources expected in our field is determined using
the Anglada et al. (1998) calculation. Assuming that extra-galactic sources have the characteristics
described by Condon (1984),   
the expected number of background sources $\langle N\rangle$ in a field of angular diameter  
$\theta_{F}$  is
\begin{equation}
\langle N\rangle=1.4\ \left[1-e^{-0.46\   
\left(\frac{S_0}{\rm mJy}\right)^{-0.75} 
\left(\frac{\theta_{F}}{500^{\prime\prime}}\right)^2  \left(\frac{\nu}{5 {\rm GHz}}\right)^2}\right]
\left(\frac{\nu}{5 {\rm GHz}}\right)^{-2.52},
\end{equation}
where  $S_0$ is the minimum detectable flux density at the centre of the field. Each of the fields has
an angular diameter of $\theta_{F}\approx500\arcsec$, so the expected number of background sources at $3.6$ cm
is
\begin{equation}
\label{eq:back1}
\langle N\rangle_{\rm 3.6cm}=2.6\ \left(\frac{S_0}{\rm 50\ \mu Jy}\right)^{-0.75}\,, 
\end{equation}
and at 6 cm,
\begin{equation}
\label{eq:back2}
\langle N\rangle_{\rm 6cm}=3.0\ \left(\frac{S_0}{\rm 100\ \mu Jy}\right)^{-0.75}\,. 
\end{equation}
Thus, we expect a total number of $\sim 9\pm3$ background sources in all our 
fields at 3.6 cm, and
$\sim 5\pm2$ sources at 6 cm. We detected 16 sources at 3.6 cm and 8 sources at 6 cm,
thus the expected number of Galactic sources detected, possibly
associated with the star forming regions observed, is $7\pm3$.

In Table~\ref{tab:radiosources} we list the detected radio sources, their
positions, flux densities and peak fluxes, 
their spectral indices $\alpha$ (in cases where a source has
been detected both in 3.6 cm and 6 cm), and their offsets from the peak of
the submm emission of the core. 
{The fluxes were corrected for the effect of primary-beam
response. We assume a detection limit of 5-$\sigma$ 
(however we make reference to a possible  4.5-$\sigma$ detection).
}
Most of the sources do not correspond to
any sources listed in the SIMBAD astronomical database and are new
sources.

In the next subsections we describe in detail the observed radio sources and
their characteristics, for each of the cores observed.

\subsection{L1582A}

\begin{figure}
\hspace{-.5cm}
\includegraphics[width=10.5cm]{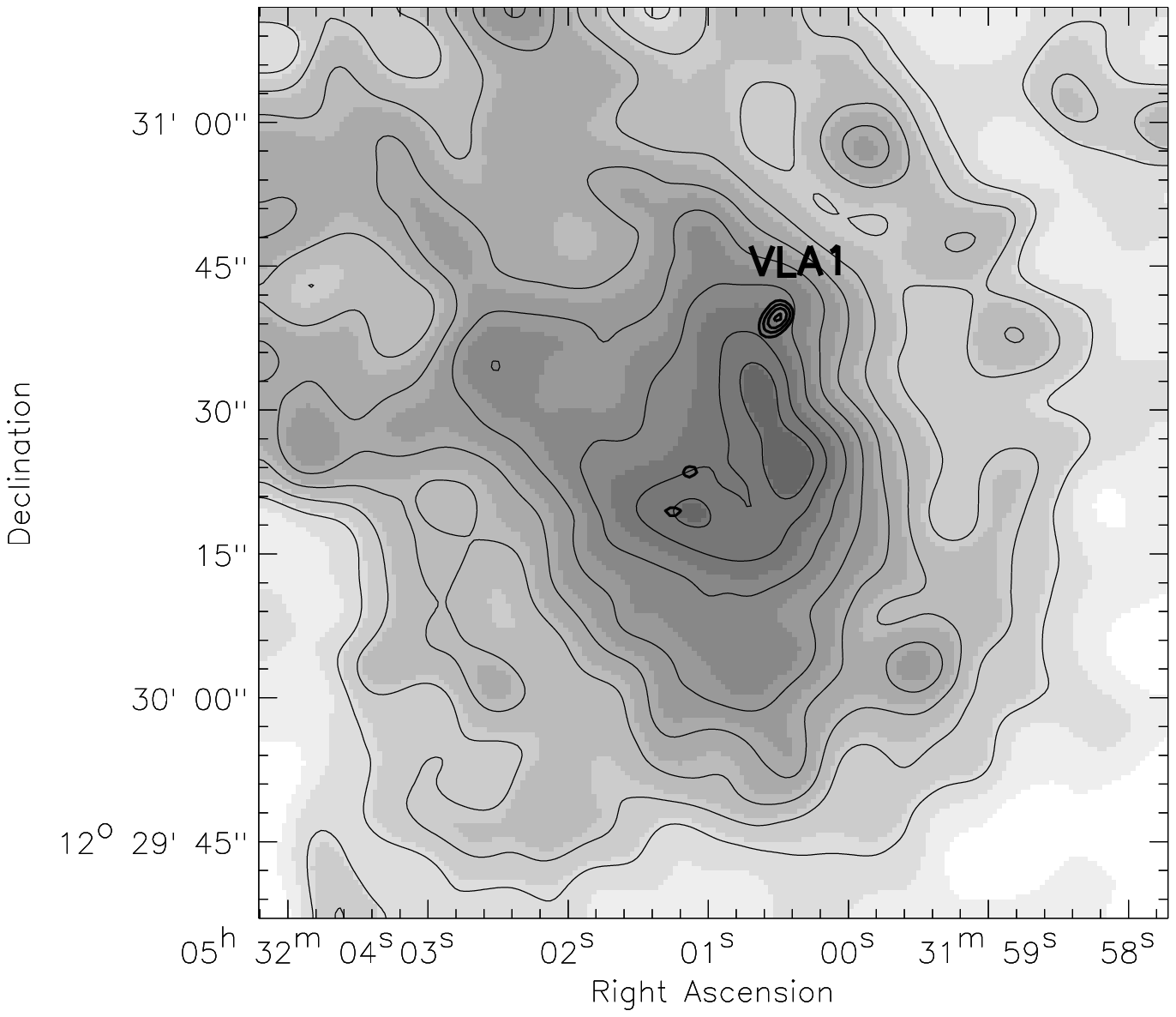}
\caption{{\bf L1582A:}  3.6 cm VLA observations (solid thick contours)
 overlaid on  850~$\micron$ SCUBA observations from Kirk et al. (2005) 
(grayscale and solid thin contours).
3.6 cm solid contours are plotted at 52 $\mu$Jy (5-$\sigma$)  $\times$ 
(1, 1.5, 2, 2.5, 3). 
850 mm dashed contours are plotted at 0.3, 0.4, 0.5, 0.6, 0.7, 0.8,
0.9 and 0.95 of the maximum 850 mm flux (170~mJy/beam).
{The VLA beam size is mentioned in Table~\ref{tab:cores}. The SCUBA beam size
at 850~$\micron$ is 14\arcsec.}
A radio source has been discovered near the centre of the core. }
\label{fig:al1582a}
\hspace{-.5cm}
\includegraphics[width=10.5cm]{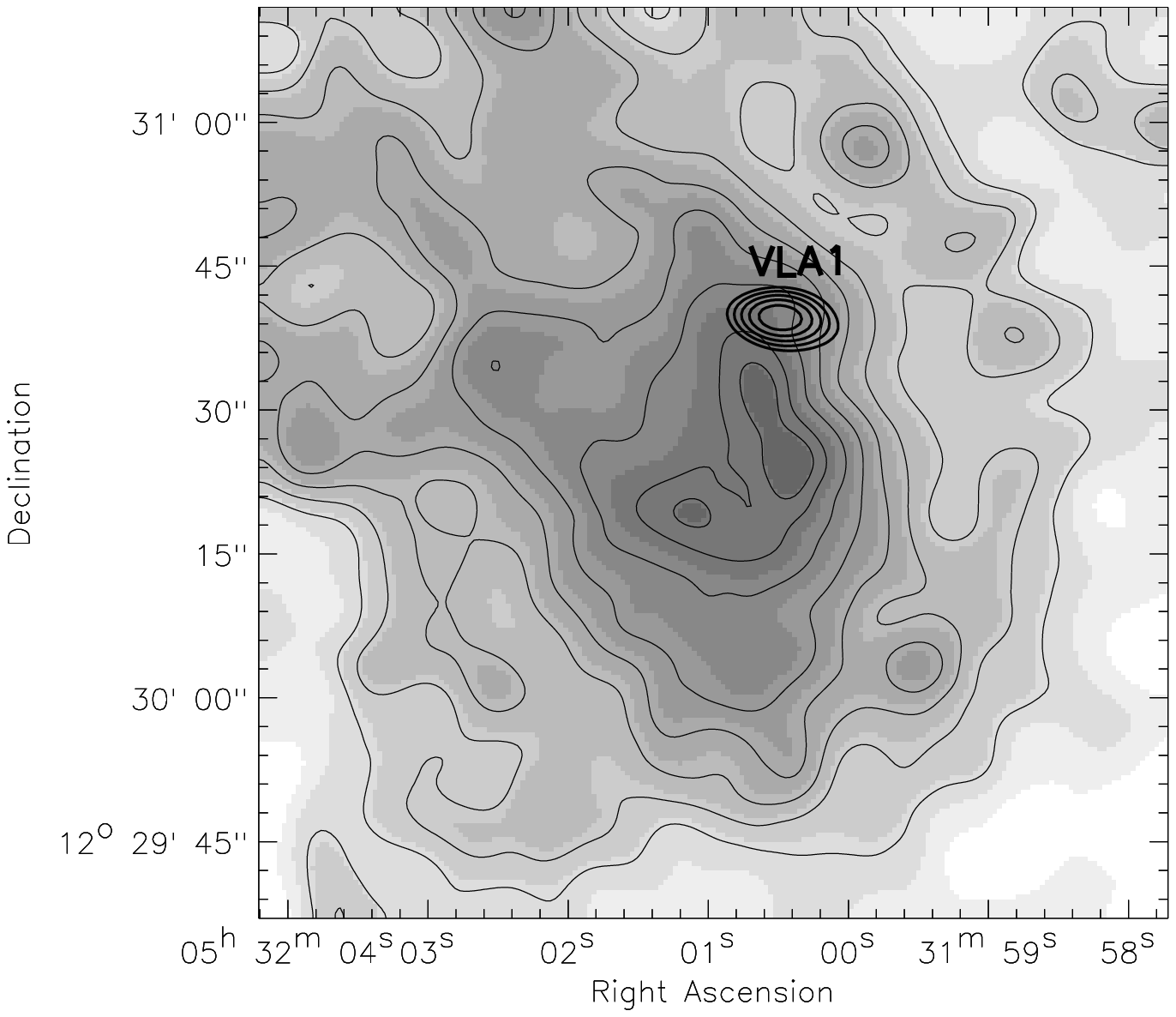}
\caption{{\bf L1582A:} 6 cm VLA observations (solid thick contours) overlaid on  
850~$\micron$ SCUBA observations from Kirk et al. (2005)
(grayscale and solid thin contours).
6 cm solid contours are plotted at 145 $\mu$Jy (5-$\sigma$)  $\times$ 
(1, 1.2, 1.4, 1.6, 1.8).
850 mm dashed contours are plotted as in Fig.~\ref{fig:al1582a}.
The VLA beam size is mentioned in Table~\ref{tab:cores}. The SCUBA beam size
at 850~$\micron$ is 14\arcsec.
}
\label{fig:bl1582a}
\end{figure}

In L1582A we discovered a weak radio source (VLA1)  close
($\sim18\arcsec$) to the peak of the submillimeter core emission. The
source is observed both in 3.6 cm and 6 cm 
(see Figs.~\ref{fig:al1582a}, \ref{fig:bl1582a}). The spectral index
of the source is $\alpha=-1.2\pm 0.5$. This spectral index is not what is
expected for Class 0 protostars. For example, for the prototypical Class 0
object VLA1623 (Andr\'e et al. 1993) $\alpha=0.6$\ . The spectral index of
the radio emission from protostars is thought to be between $-0.1$ and
$2$, consistent with  thermal free-free emission, whereas
for extra-galactic sources it is expected to be $<-0.1$, characteristic of
non-thermal emission due to optically thin synchrotron emission (e.g.
Andr\'e et al. 1987; Anglada et al. 1998; Beltran et al. 2001).  Thus, 
L1582A-VLA1 is most probably an extragalactic source.

However we  note that recent observations of the confirmed
low luminosity protostar L1014-IRS (Shirley et al., VLA proposal)
indicate a similar spectral index.   The unusual negative
spectral index  may be due to non-thermal (synchrotron) emission.
The similarities of L1582A-VLA1 to 
L1014-IRS (similar 3.6 cm and 6 cm fluxes, similar offsets from the peak
of the submillimeter emission of the core) are suggestive that L1582A-VLA1 
may also be a very young protostar embedded in the core, but this cannot be 
confirmed with the current observations.

Two more sources were detected in the field. VLA2 was detected only at 3.6
cm but a negative spectral index cannot be excluded. This source is
outside the core but within the extended CO ($J=1-0$) line emission (Dame
et al. 2001; see also Kirk et al. 2005). VLA3 was detected only at 6cm and
thus has a negative spectral index and should be a background source.

Thus, we are unable to confirm unequivocally that L1582A is starless.

\subsection{L1689A}

In the L1689A field we discovered 6 sources, 5 of which were detected both
at 3.6 cm and 6 cm, and 1 marginally detected only at 3.6 cm.

\begin{figure}[!b]
\hspace{-1.2cm}
\includegraphics[height=8.3cm]{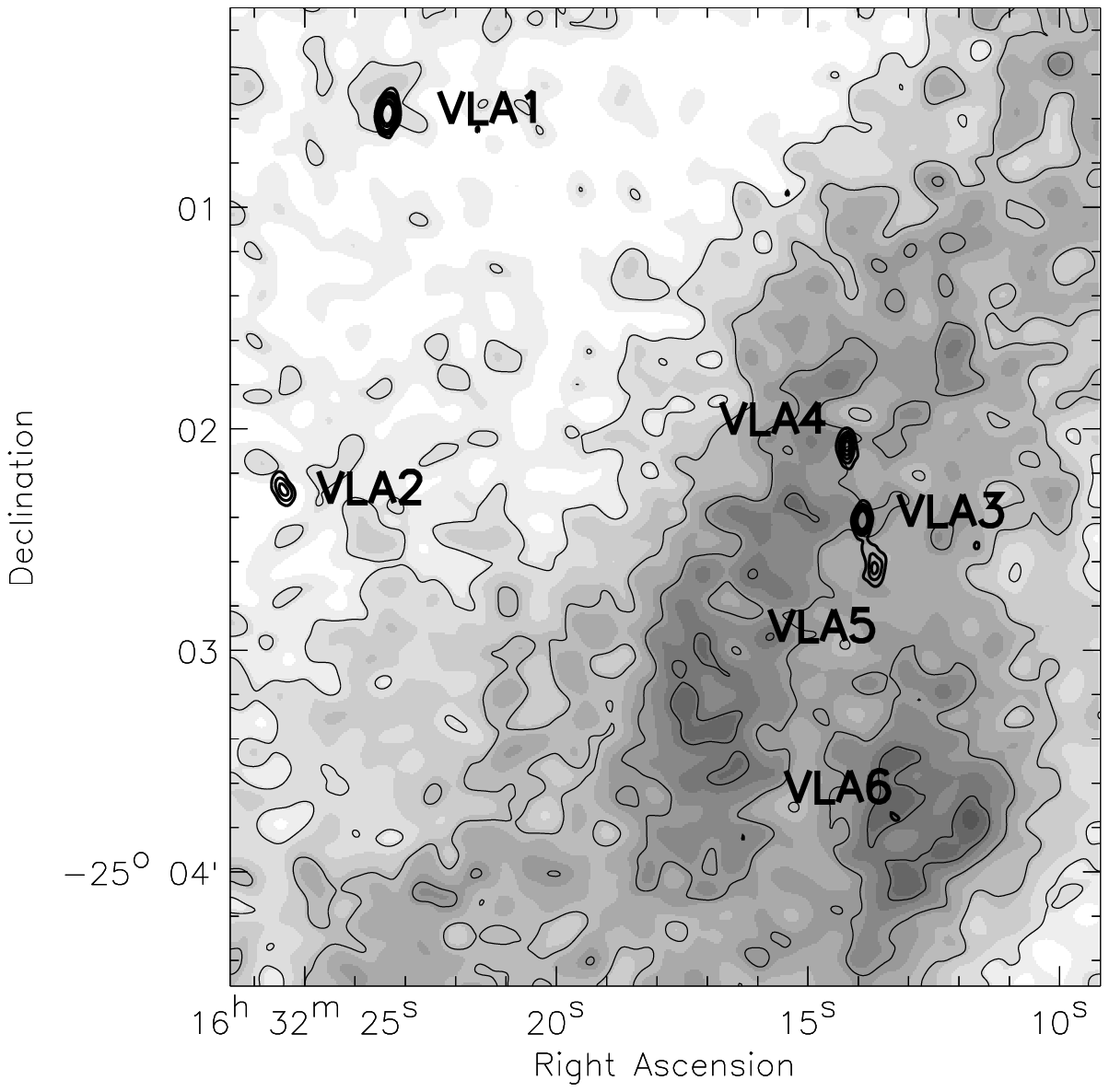}
\caption{{\bf L1689A:}  3.6 cm VLA observations (solid thick contours)
 overlaid on  850~$\micron$ SCUBA observations from Nutter et al. (2006) 
(grayscale and thin solid contours).
3.6 cm solid contours are plotted at 54 $\mu$Jy (4.5-$\sigma$)  $\times$ 
(1, 2, 4, 6, 8, 10, 12). 
850 mm dashed contours are plotted at 0.2, 0.4, 0.6, and
0.8  of the maximum 850 mm flux (170~mJy/beam).The VLA beam size is mentioned in 
Table~\ref{tab:cores}. The SCUBA beam size
at 850~$\micron$ is 14\arcsec.
}
\label{fig:al1689a}
\hspace{-1.2cm}
\includegraphics[height=8.3cm]{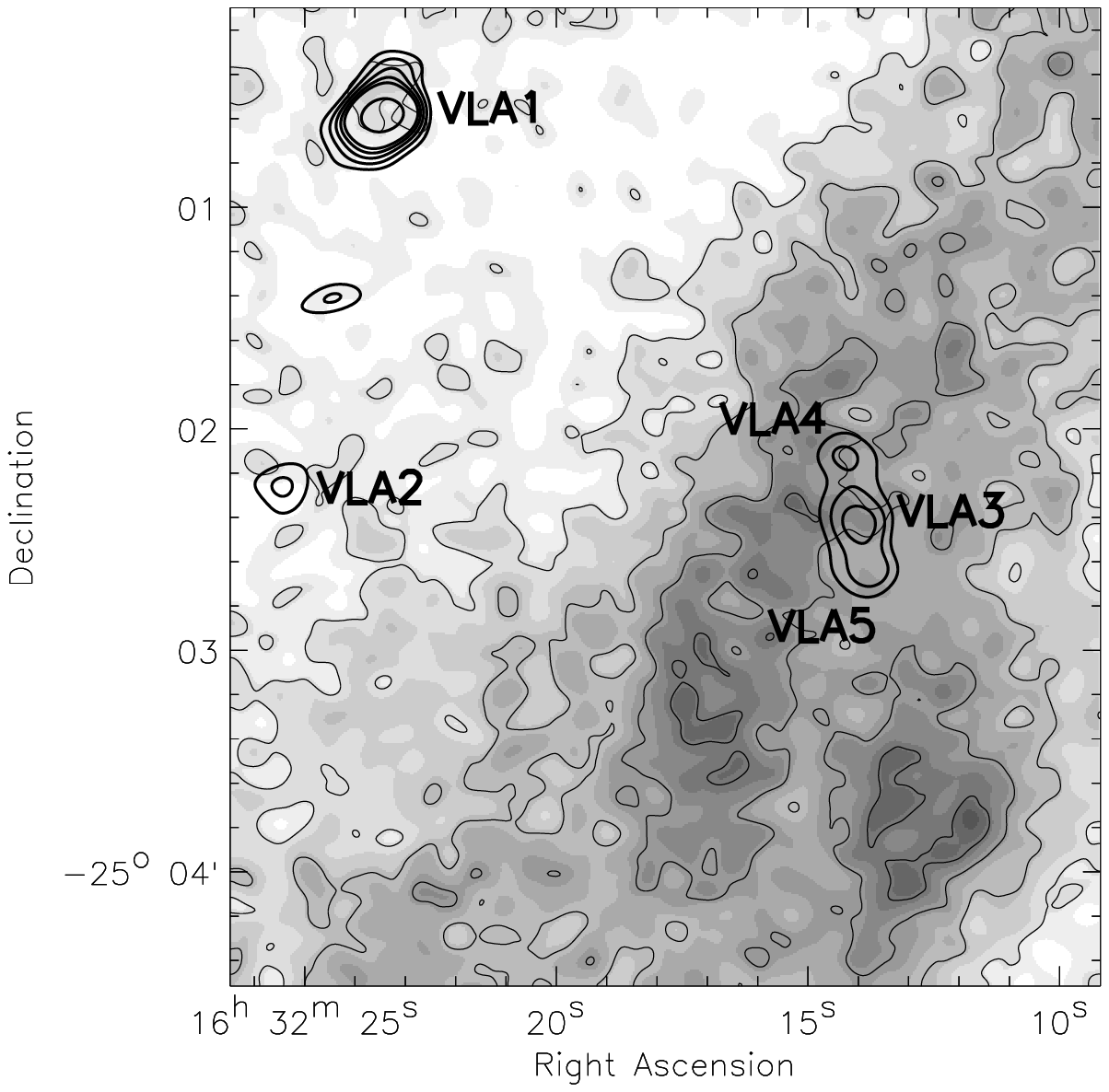}
\caption{{\bf L1689A:}  
 6 cm VLA observations (solid thick contours)
 overlaid on  850~$\micron$ SCUBA observations from Nutter et al. (2006) 
(grayscale and thin solid contours).
6 cm solid contours are plotted at 300 $\mu$Jy (5-$\sigma$)  $\times$ 
(1, 1.5, 2, 2.5, 3, 6, 9). 
850 mm dashed contours are plotted as in Fig.~\ref{fig:al1689a}.
The VLA beam size is mentioned in Table~\ref{tab:cores}. The SCUBA beam size
at 850~$\micron$ is 14\arcsec.}
\label{fig:bl1689a}
\end{figure}

VLA1 has a positive spectral index, indicative of free-free thermal
emission from ionised gas. It is outside the main core but within the
extended CO line emission of the Ophiuchus GMC (Dame et al. 2001; Kirk et
al. 2005). It coincides with a local column density peak of the cloud (see
Fig.~\ref{fig:al1689a} and Nutter et al. 2006), although
the dust emission is rather weak. 

VLA2 is also outside the core but within the extended CO emission. 
The sign of its spectral index cannot be determined; it could be either a 
young star or a background source. 

Sources VLA3, VLA4 and VLA5 were detected in both wavelengths and they
seem to be related, as they are aligned. They are located outside the
L1689A core but they are coincident with a submm filament adjacent to the
core. The total projected extent of the structure is $\sim 46\arcsec$
(i.e. $\sim$ 6000~AU if they are physically located close to L1689A). The
central (VLA3) and the southern (VLA5) source have negative spectral
indices (characteristic of non-thermal synchrotron emission), 
whereas the northern source (VLA4) may have positive or marginally negative
spectral index (characteristic of optically thin thermal emission).  
Thus, these sources are probably extragalactic.

The last source in this field, VLA6, was marginally detected at 3.6 cm
(4.5-$\sigma$ detection).  It is located at the centre of the core. It was
not detected at 6 cm but a negative spectral index cannot be ruled out.
The presence of this source in the centre of the core, if indeed it is
real, would explain the temperature increase toward the centre of the core
suggested by the observations of Ward-Thompson et al. (2002). Thus, there
is the possibility that L1689A may not be starless.

\subsection{B133}

In the B133 field we discovered 4 radio sources. The strongest one,
B133-VLA1, is located outside the core. It has a negative spectral index
between 3.6 and 6 cm, indicating  non-thermal emission due to optically
thin synchrotron emission characteristic from an extra-galactic source. This
source was the only one from this field that was detected at 6 cm. A
search in the SIMBAD database reveals a match with PMNJ1906-0650 of the
Parkes-MIT-NRAO (PMN) Surveys (Griffith et al. 1995) at 4.85GHz
($S_\nu=65\pm11$ mJy) . The source has been detected by the NRAO VLA Sky
Survey (Condon et al. 1998)  at 1.4 GHz ($S_\nu=295.5\pm8.9$ mJy), and by
the Texas Survey of Discrete Radio Sources (Douglas et al. 1996) at 365
MHz ($S_\nu=1216\pm41$ mJy). These fluxes are also consistent with thin
synchrotron emission from an extra-galactic source.

\begin{figure}[!h]
\hspace{-0.9cm}
\includegraphics[height=8.3cm]{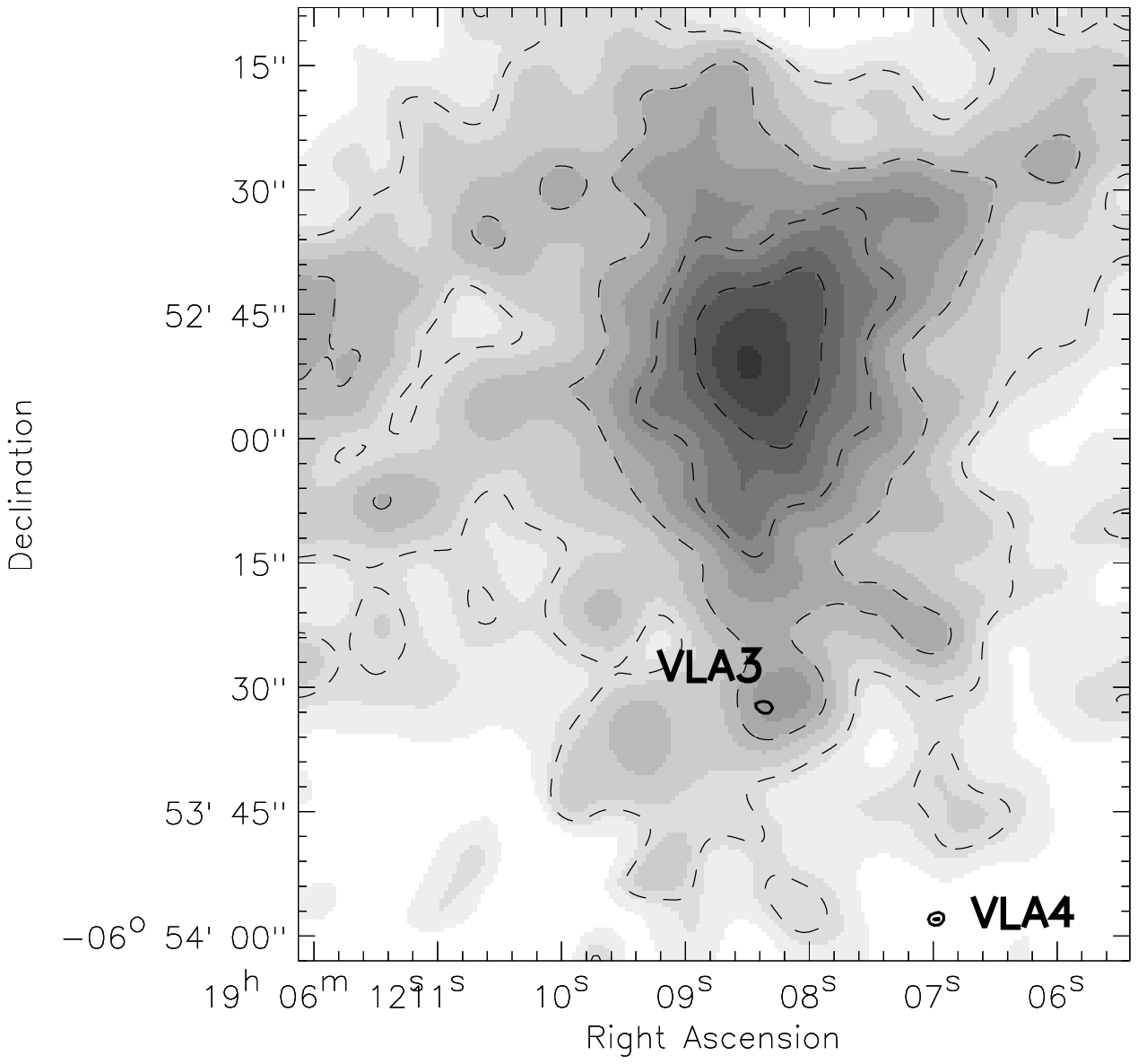}
\caption{{\bf B133:}  
 3.6 cm VLA observations (solid contours)
 overlaid on  850~$\micron$ SCUBA observations from Kirk et al. (2006) 
(grayscale and dashed contours).
3.6 cm solid contours are plotted at 60 $\mu$Jy (5-$\sigma$)  $\times$ 
(1, 1.1, 1.2, 1.3). 
850 mm dashed contours are plotted at 0.2, 0.4, 0.6, and 0.8
of the maximum 850 mm flux (170~mJy/beam). The VLA beam size is mentioned in 
Table~\ref{tab:cores}. The SCUBA beam size is 14\arcsec.
}
\label{fig:ab133}
\end{figure}

The second strongest source in the field, B133-VLA2, is located outside
the core and it is not associated with any column density peak.  The
absence of detectable 6 cm emission at the 300~$\mu$Jy level indicates a
positive spectral index ($\alpha>1.7\pm0.2$).  This suggests free-free
thermal emission. B133-VLA2 may be 
associated with the Aquila Rift, where B133 is thought to reside
(Kirk et al. 2005).

The B133-VLA3 source is located inside the core within a local column density
peak.  It has been marginally detected (at the $\sim$ 6-$\sigma$ level) and
it could be a very young protostar embedded in the core. Unfortunately,
although it is not detected at 6 cm a negative spectral index cannot be
excluded, so it may also be an extragalactic source.

Finally, B133-VLA4 is marginally detected at 3.6 cm. It is not detected at
6 cm but a negative spectral index is possible.

Thus, based on the VLA observations B133 appears to be a starless core with a
possible protostar (B133-VLA3) at its edge.

\subsection{B68}

Four radio sources were discovered in the B68 field at 3.6 cm.  All of
them are outside the core and they are not associated with the core but
they are within the extended CO emission (Dame et al. 2001; Kirk et 2005)
around the core. This core was not observed at 6 cm, and thus the spectral
indices of these sources and their nature cannot be determined.  We
conclude that B68 is starless.

\section{Summary}
 
We conducted continuum VLA radio observations of 4 dense cores previously
classified as prestellar, searching for radio emission from deeply
embedded protostars. 17 radio sources were detected in the observed
fields, i.e. more than the expected number of background sources. Thus at
least a few of them may be associated with the star forming regions
observed.

Two radio sources were discovered near the centres of the cores.
L1582A-VLA1 is located within 18\arcsec of the peak of the submm emission
of the core, and its similarity with the source embedded in L1014
suggests that this may also be a very young protostar embedded in the core,
despite its  negative spectral index between 3.6 and 6cm, and its offset from the
centre of the core.   
L1689A-VLA6 is located within 1\arcsec of the peak of the submm emission,
and it may also be a very young protostar embedded in the core.  
However, this detection was only at the 4.5-$\sigma$ level.  
Future {\it Spitzer} data should confirm whether these are young protostars or not.
{If these two sources  indeed are associated with the observed cores then they are similar to
the recently discovered low-luminosity protostars, such as that in L1521F, which
are young Class 0 objects, e.g. like VLA1623 (Andr\'e et al. 1993)
and  HH24MMS (Ward-Thompson et al. 1995; Bontemps et al. 1996).
However, based on the current observations and the locations of the
radio sources, we conclude that 
L1582A and L1689A are most likely starless.}

Two radio sources were discovered close to 
edges of cores. L1689A-VLA345, at the edge of the L1689A, 
 is probably an extragalactic source. B133-VLA3, at the edge of B133, 
coincides with a local peak in the core submillimeter emission, and it may be 
a young protostar.

Four radio sources with positive spectral indices were detected outside
the cores (L1689A-VLA1, L1689A-VLA2, B133-VLA2, B133-VLA4) and, thus could
also be young protostars or stars associated with the extended star forming regions
where these cores reside.

All of the candidate protostars reported are expected to be detectable in
the NIR by the {\it Spitzer} space telescope and such observations are needed to
confirm whether these sources are indeed protostars.  Additionally,
sensitive molecular line radio observations, with e.g. the Submillimeter
Array (SMA) should be able to detect molecular outflows emanating from
these objects, if they are protostars.

\section{Conclusion}

In this paper we presented deep radio observations of 4 prestellar cores, which
are relatively warm and almost circular (in their inner regions). 
 According to the Stamatellos et al. (2005) 
criteria these cores are, among cores presently classified as prestellar, the 
ones most likely to
nurture very young protostars. {However, we find no definite protostars at the centres
of these cores. Two radio sources that
could be very young embedded 
protostars were discovered near the centres of L1582A and L1689A, but further
near- or mid-infrared observations 
(e.g. with {\it Spitzer} or the upcoming {\it Herschel})
are needed for safe conclusions on the nature of
these sources.}
Hence, based on the current observations, similarly to Kirk et al. (2006)
we conclude that the number of cores misclassified as prestellar
is probably very small and does not affect significantly the estimated lifetime of the 
prestellar phase. 

\begin{acknowledgements}

We thank C. Chandler for her help in reducing and calibrating the VLA
data at NRAO Sorocco, and J. Kirk and D. Nutter for
providing the SCUBA data. The James Clerk Maxwell Telescope is operated by
The Joint Astronomy Centre on behalf of the Particle Physics and Astronomy
Research Council of the United Kingdom, the Netherlands Organisation for
Scientific Research, and the National Research Council of Canada. The JCMT
data shown in this paper were taken during observing runs M96BU56,
M97BU88, M99AU34 and M00AU32. This work was carried out while DWT was on sabbatical
at l'Observatoire de Bordeaux and CEA, Saclay, and he wishes to thank both
institutions for the hospitality accorded to him.
We also acknowledge support by PPARC grant PPA/G/O/2002/00497.

\end{acknowledgements}


\begin{thebibliography}{}


\bibitem[Andre et al.(1987)]{1987AJ.....93.1182A} Andr\'e, P., Montmerle, T., 
\& Feigelson, E.~D.\ 1987, \aj, 93, 1182 

\bibitem[Andre et al.(1988)]{1988ApJ...335..940A} Andr\'e, P., Montmerle, T., 
Feigelson, E.~D., Stine, P.~C., \& Klein, K.-L.\ 1988, \apj, 335, 940 

\bibitem[1993]{andre2} Andr\'e, P., Ward-Thompson, D., \& Barsony, M.\ 1993, ApJ, 406, 122 

\bibitem[Andre(1996)]{1996ASPC...93..273A} Andre, P.\ 1996, ASP Conf.~Ser.~ 
93: Radio Emission from the Stars and the Sun, 93, 273 

\bibitem[Andr\'e, Ward-Thompson, \& Barsony(2000)] 
{2000prpl.conf...59A} 
Andr\'e, P., Ward-Thompson, D., \& Barsony, M.\ 2000, 
Protostars and Planets IV, 59 

 
\bibitem[1998]{anglada} Anglada, G., Villuendas, E., Estalella, R., Beltr{\'a}n, M.~T. et al.
 \ 1998, AJ, 116, 2953 


\bibitem[2001]{beltran} Beltr{\'a}n, M.~T., 
Estalella, R., Anglada, G., Rodr{\'{\i}}guez, L.~F., \& Torrelles, J.~M.\ 
2001, AJ, 121, 1556 

\bibitem[Bontemps, Andre, \& Ward-Thompson(1995)]{1995A&A...297...98B} 
Bontemps, S., Andr\'e, P., \& Ward-Thompson, D.\ 1995, \aap, 297, 98

\bibitem[Bontemps et al.(1996)]{1996A&A...314..477B} Bontemps, S., 
Ward-Thompson, D., \& Andre, P.\ 1996, \aap, 314, 477 

\bibitem[Bourke et al.(2005)]{2005ApJ...633L.129B} Bourke, T.~L., Crapsi, 
A., Myers, P.~C., Evans, N.~J., Wilner, D.~J., Huard, T.~L., J{\o}rgensen, 
J.~K., \& Young, C.~H.\ 2005, \apjl, 633, L129 

\bibitem[Bourke et al.(2006)]{2006ApJ...649L..37B} Bourke, T.~L., et al.\ 
2006, \apjl, 649, L37 

\bibitem[Cassen \& Moosman(1981)]{1981Icar...48..353C} Cassen, P.~\& 
Moosman, A.\ 1981, Icarus, 48, 353 

\bibitem[Condon(1984)]{1984ApJ...287..461C} Condon, J.~J.\ 1984, \apj, 287, 
461 

\bibitem[Condon et al.(1998)]{1998AJ....115.1693C} Condon, J.~J., Cotton, 
W.~D., Greisen, E.~W., Yin, Q.~F., Perley, R.~A., Taylor, G.~B., \& 
Broderick, J.~J.\ 1998, \aj, 115, 1693 

\bibitem[Curiel, Canto, \& Rodr\'iguez(1987)]{1987RMxAA..14..595C} Curiel, 
S., Canto, J., \& Rodr\'iguez, L.~F.\ 1987, Revista Mexicana de Astronomia y 
Astrofisica, vol.~14, 595

\bibitem[Curiel et al.(1993)]{1993ApJ...415..191C} Curiel, S., Rodr\'iguez, 
L.~F., Moran, J.~M., \& Canto, J.\ 1993, \apj, 415, 191 


\bibitem[Curiel(1995)]{1995RMxAC...1...59C} Curiel, S.\ 1995, Revista 
Mexicana de Astronomia y Astrofisica Conference Series, 1, 59 

\bibitem[Dame et al.(2001)]{2001ApJ...547..792D} Dame, T.~M., Hartmann, D., 
\& Thaddeus, P.\ 2001, \apj, 547, 792 

\bibitem[Douglas et al.(1996)]{1996AJ....111.1945D} Douglas, J.~N., Bash, 
F.~N., Bozyan, F.~A., Torrence, G.~W., \& Wolfe, C.\ 1996, \aj, 111, 1945 

\bibitem[Eiroa et al.(2005)]{2005AJ....130..643E} Eiroa, C., Torrelles, 
J.~M., Curiel, S., \& Djupvik, A.~A.\ 2005, \aj, 130, 643 

\bibitem[Feigelson et al.(1998)]{1998ApJ...494L.215F} Feigelson, E.~D., 
Carkner, L., \& Wilking, B.~A.\ 1998, \apjl, 494, L215 



\bibitem[Griffith et al.(1995)]{1995ApJS...97..347G} Griffith, M.~R., 
Wright, A.~E., Burke, B.~F., \& Ekers, R.~D.\ 1995, \apjs, 97, 347 

\bibitem[Goodwin et al.(2002)]{2002MNRAS.330..769G} Goodwin, S.~P., 
Ward-Thompson, D., \& Whitworth, A.~P.\ 2002, \mnras, 330, 769 

\bibitem[Harvey et al.(2002)]{2002AJ....123.3325H} Harvey, D.~W.~A., 
Wilner, D.~J., Di Francesco, J., Lee, C.~W., Myers, P.~C., \& Williams, 
J.~P.\ 2002, \aj, 123, 3325 

\bibitem[Huard et al.(2006)]{2006ApJ...640..391H} Huard, T.~L., et al.\ 
2006, \apj, 640, 391 

\bibitem[Kauffmann et al.(2005)]{2005AN....326..878K} Kauffmann, J., 
Bertoldi, F., Evans, N.~J., \& the C2D Collaboration 2005, Astronomische 
Nachrichten, 326, 878 

\bibitem[kirk]{2005} Kirk, J.~M., 
Ward-Thompson, D., \& Andr{\'e}, P.\ 2005, MNRAS, 360, 1506 

\bibitem[kirk]{2006} Kirk, J.~M., 
Ward-Thompson, D., \& Andr{\'e}, P.\ 2006, MNRAS submitted

\bibitem[Leous et al.(1991)]{1991ApJ...379..683L} Leous, J.~A., Feigelson, 
E.~D., Andr\'e, P., \& Montmerle, T.\ 1991, \apj, 379, 683 

\bibitem[Martin(1996)]{1996ApJ...473.1051M} Martin, S.~C.\ 1996, \apj, 473, 
1051 

\bibitem[1983]{myers} Myers, P.~C., Linke, R.~A., \& Benson, P.~J.\ 1983, ApJ, 264, 517 

\bibitem[Neufeld \& Hollenbach(1994)]{1994ApJ...428..170N} Neufeld, 
D.~A.~\& Hollenbach, D.~J.\ 1994, \apj, 428, 170 

\bibitem[Neufeld \& Hollenbach(1996)]{1996ApJ...471L..45N} Neufeld, 
D.~A.~\& Hollenbach, D.~J.\ 1996, \apjl, 471, L45 

\bibitem[Nutter et al.(2006)]{2006MNRAS.368.1833N} Nutter, D., 
Ward-Thompson, D., \& Andr{\'e}, P.\ 2006, \mnras, 368, 1833 

\bibitem[Raga et al.(2000)]{2000A&A...364..763R} Raga, A.~C., Curiel, S., 
Rodr{\'{\i}}guez, L.~F., \& Cant{\'o}, J.\ 2000, \aap, 364, 763 

\bibitem[Rodriguez et al.(1989)]{1989ApJ...346L..85R} Rodr\'iguez, L.~F., 
Curiel, S., Moran, J.~M., Mirabel, I.~F., Roth, M., \& Garay, G.\ 1989a, 
\apjl, 346, L85

\bibitem[Rodriguez et al.(1989)]{1989ApJ...347..461R} Rodr\'iguez, L.~F., 
Myers, P.~C., Cruz-Gonzalez, I., \& Terebey, S.\ 1989b, \apj, 347, 461 
 
\bibitem[Rodr{\'{\i}}guez et al.(1999)]{1999ApJS..125..427R} 
Rodr{\'{\i}}guez, L.~F., Anglada, G., \& Curiel, S.\ 1999, \apjs, 125, 427 
 
\bibitem[2005]{stama} Stamatellos, D., Whitworth, A.P., Boyd, D.F.A.,\& Goodwin, S.P. \ 2005a, 
A\&A, 439, 159

\bibitem[Stamatellos et al.(2005)]{2005AN....326..882S} Stamatellos, D., 
Whitworth, A.~P., \& Goodwin, S.~P.\ 2005b, Astronomische Nachrichten, 326, 
882 

\bibitem[Terebey et al.(2005)]{2005prpl.conf.8611T} Terebey, S., van Buren, 
D., Brundage, M., \& Hancock, T.\ 2005, Protostars and Planets V, 
Proceedings of the Conference held October 24-28, 2005, in Hilton Waikoloa 
Village, Hawai'i.~LPI Contribution No.~1286., p.8611 

\bibitem[1994]{derek} Ward-Thompson, D., Scott, P.~F., Hills, R.~E., \& Andr\'e, P.\ 1994,
 MNRAS, 268, 276 

\bibitem[Ward-Thompson et al.(1995)]{1995MNRAS.274.1219W} Ward-Thompson, 
D., Chini, R., Krugel, E., Andre, P., \& Bontemps, S.\ 1995, \mnras, 274, 
1219 

\bibitem[Ward-Thompson, Andr{\' e}, \& Kirk(2002)] 
{2002MNRAS.329..257W} 
Ward-Thompson, D., Andr{\' e}, P., \& Kirk, J.~M.\ 2002, \mnras, 329, 257

\bibitem[Winkler \& Newman(1980)]{1980ApJ...236..201W} Winkler, K.-H.~A.~\& 
Newman, M.~J.\ 1980, \apj, 236, 201 

\bibitem[2004]{young} Young, C.~H., et al.\ 2004, ApJS, 154, 396 


\end{thebibliography}
\end{document}